\documentclass[
    ,final            % use final for the camera ready runs
  ]
  {aipproc}
\layoutstyle{8x11single}

\begin{document}

\title{Two-color QCD at high density}

%\classification{<Replace this text with PACS numbers; choose from this list:
%                \texttt{http://www.aip..org/pacs/index.html}>}

\author{Tamer Boz}{
  address={Department of Mathematical Physics, Maynooth University, Maynooth, Co. Kildare, Ireland}
    ,altaddress={Centre for the Subatomic Structure of Matter, Adelaide University, Adelaide, SA 5005, Australia} % additional visiting address
}

\author{Pietro Giudice}{
  address={Universit{\"a}t M{\"u}nster, Institut f{\"u}r Theoretische Physik, M{\"u}nster, Germany}
}

\author{Simon Hands}{
  address={Department of Physics, College of Science, Swansea University, Swansea, United
Kingdom}
}

\author{Jon-Ivar Skullerud}{
  address={Department of Mathematical Physics, Maynooth University, Maynooth, Co. Kildare, Ireland}
  ,altaddress={Centre for the Subatomic Structure of Matter, Adelaide University, Adelaide, SA 5005, Australia} % additional visiting address
}

\author{\newline Anthony G. Williams}{
  address={Centre for the Subatomic Structure of Matter, Adelaide University, Adelaide, SA 5005, Australia}
}

\begin{abstract}
QCD at high chemical potential has interesting properties such as deconfinement of quarks. Two-color QCD, which enables numerical simulations on the lattice, constitutes a laboratory to study QCD at high chemical potential. Among the interesting properties of two-color QCD at high density is the diquark condensation, for which we present recent results obtained on a finer lattice compared to previous studies. The quark propagator in two-color QCD at non-zero chemical potential is referred to as the Gor'kov propagator. We express the Gor'kov propagator in terms of form factors and present recent lattice simulation results.
\end{abstract}

\maketitle

%%%%%%%%%%%%%%%%%%%%%%%%%%%%%%%%%%%%%%%%%%%%
%% MAINMATTER
%%%%%%%%%%%%%%%%%%%%%%%%%%%%%%%%%%%%%%%%%%%%

\section{Introduction}

At extremely high chemical potential, real QCD (three-color QCD) is expected to exhibit interesting phenomena like deconfinement of quarks or a QCD-analog of the superconducting phase of QED, and these may exist in compact stars \cite{Alford2008}. It is therefore desirable to investigate QCD in this region. However there is the so-called sign problem for real QCD which disables the use of Monte-Carlo simulations.
At this point two-color QCD ($N_{c}=2$) with an even number of flavors ($N_{f}$) comes to our aid: it turns out that the sign problem disappears for this theory. This means that we have a tool to attack QCD from first principles at high density, the only problem being that the theory is not physical. Fortunately, at low baryon density two-color QCD still has the properties such as chiral symmetry breaking and confinement. A nice review of QCD at high density has been given in \cite{Muroya2003}.

Two-color QCD has been studied by several groups \cite{Hands2000, Hands2001, Kogut2001, Hands2002, Hands2006, Hands2007, Hands2010, Cotter2013, Boz2013}, and these studies led to a tentative phase diagram of the theory.
Figure 1 shows a tentative phase diagram of two-color QCD with $N_{f}=2$ taken from \cite{Boz2013}. The bare lattice parameters are $ \kappa=0.168 $ and $\beta=1.9$ corresponding to a lattice spacing of $a\approx0.18$fm. The corresponding pion-to-rho meson mass ratio is $m_{\pi}/m_{\rho}=0.8$. The green area denotes the crossover region. The area above the crossover region is a quark-gluon plasma. At low temperature and low chemical potential there is a hadronic phase. For $ N_{f}=N_{c}=2 $, the hadrons of the theory are mesons and baryons, which are equivalent. The circles and diamonds denote pseudocritical points for superfluid to normal and deconfinement transitions, respectively. We see that the critical temperature for the superfluid to normal transition does not depend on the chemical potential once the chemical potential is above the onset transition. The same thing is not true for the deconfinement transition, as the blue curve suggests a decrease in the critical temperature with an increase in the chemical potential.

\begin{figure}[h!]
\includegraphics[width=12.6cm,height=9.0cm]{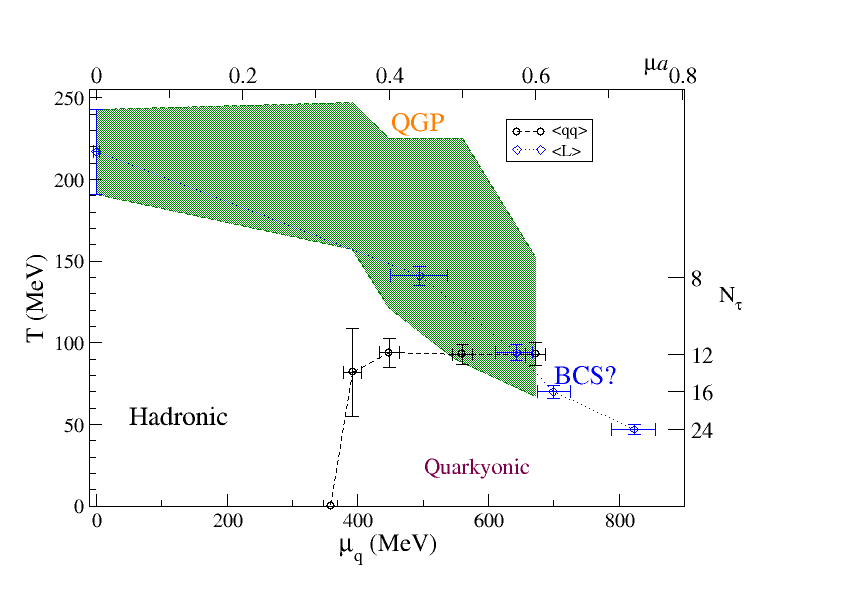}
\caption{Tentative phase diagram of two-color QCD, taken from \cite{Boz2013}.}
\end{figure}

\section{Simulation details}

The action of the theory we consider in this work is
\begin{equation}
S=\overline{\psi}_{1}M\left(\mu\right)\psi_{1}+\overline{\psi}_{2}M\left(\mu\right)\psi_{2}-J\overline{\psi}_{1}\left(C\gamma_{5}\right)\tau_{2}\overline{\psi}_{2}^{tr}
+\overline{J}\psi_{2}^{tr}\left(C\gamma_{5}\right)\tau_{2}\psi_{1},\label{eq.1}
\end{equation}
where $\psi_{1}$ and $\psi_{2}$ denote the two flavor fields, $M$ is the Wilson fermion matrix and $C$ is the charge conjugation operator.
The Wilson fermion matrix in the presence of a chemical potential, $\mu\neq0$, and in the absence of diquark source, $j=0$, in position space is given by:
\begin{equation}
M(\mu)=\delta_{xy}-\kappa\sum_{\nu}\left[\left(\mathbf{1}-\gamma_{\nu}\right)\mbox{e}^{\mu\delta_{\nu0}}U_{\nu}\left(x\right)\delta_{y,x+\hat{\nu}}+\left(\mathbf{1}+\gamma_{\nu}\right)\mbox{e}^{-\mu\delta_{\nu0}}U_{\nu}^{\dagger}\left(y\right)\delta_{y,x-\hat{\nu}}\right].
\end{equation}
Here $U\left(x\right)$ is the gauge field, and $U\left(x\right)\neq0$ means that the fermion matrix expression is for the case when there is interaction. The Fourier transform of $M\left(\mu\right)$ gives the Wilson fermion matrix in the momentum space (for $U\left(x\right)=\mathbf{1}$):

\begin{equation}
M\left(p\right)=\mbox{i}\sum_{j=1}^{3}\gamma_{j}\mbox{sin}\left(p_{j}\right)+\mbox{i}\gamma_{4}\mbox{sin}\left(\omega\right)+m_{0}+\sum_{j=1}^{3}\left[\mathbf{1}-\mbox{cos}\left(p_{j}\right)\right]+\left[\mathbf{1}-\mbox{cos}\left(\omega\right)\right],
\label{eq.3}
\end{equation}
where $\omega=p_{4}-\mbox{i}\mu$. 

The last two terms in (\ref{eq.1}) are introduced in order to calculate the  \textit{diquark condensate}, which is explained in Section 3. The factor $J$ includes the \textit{diquark source}, $j$, which controls these terms and is sent to zero to obtain the physical limit. Such a calculation is carried out with the parameters $\beta=2.1$, $\kappa=0.1577$, $a\approx0.125$ fm, and for two different diquark sources, $ja=0.02$ and $ja=0.03$ (The corresponding pion-to-rho meson mass ratio is $m_{\pi}/m_{\rho}=0.8$). The quark number density is also calculated with the same parameters. The simulation parameters used to obtain the results for the form factors of the Gor'kov propagator, which is explained in Section 4, are $\beta=1.9$,  $\kappa=0.1680$, $a\approx0.18 $ fm and $ja=0.04$.

In the calculation of the form factors of the quark propagator, we have fixed the configurations to Landau gauge.

\section{Diquark condensation in two-color QCD with $N_{f}=2$}

In a theory with $ N_{c}=2 $, quarks and antiquarks live in equivalent representations of the color group. Baryons of this theory are diquarks, and at zero chemical potential there is an exact symmetry between the diquarks and mesons. At zero chemical potential, the pseudo-Goldstone multiplet consists of the pion isotriplet plus a scalar isoscalar diquark and antidiquark.

When the chemical potential reaches the value of about half the mass of a pion, these diquark baryons are expected to condense. This is called the diquark condensate, and this condensation gives rise to a superfluid phase. In real QCD, the analog of such a phase is a superconducting phase because these diquarks are not gauge invariant in the world of three colors.

The order parameter of the transition to this superfluid phase is the diquark condensate:

\begin{equation}
\left\langle qq\right\rangle \equiv\left\langle q^{T}C\gamma_{5}\tau_{2}q\right\rangle,
\end{equation}
where $C$ is the charge conjugation operator.

Figure 2 shows a plot of the diquark condensate calculated in this way, with respect to the chemical potential. The physical curve here is the curve for $j=0$, which is linearly extrapolated from the data for $ja=0.02$ and $ja=0.03$. The diquark condensation takes place when the chemical potential reaches the value of about half the pion mass, which means that there is enough energy to excite diquarks and give rise to the superfluid phase. 

The parameters of the calculation correspond to a value of $m_{\pi}a=0.446(3) $ for the pion mass. Therefore a transition at about $\mu a=0.223$ is expected, and the curve for $ja=0$ is consistent with this expectation. We note that the extrapolation to $j=0$ is poor in that the curve does not cross zero at $\mu a=0.2$. This is due to the fact that linear extrapolation was used with diquark condensate values corresponding to only two different diquark sources. We plan to improve this result by adding diquark condensate values for yet another diquark source, $ja=0.01$.

\begin{figure}
\includegraphics[width=12.6cm,height=9.0cm]{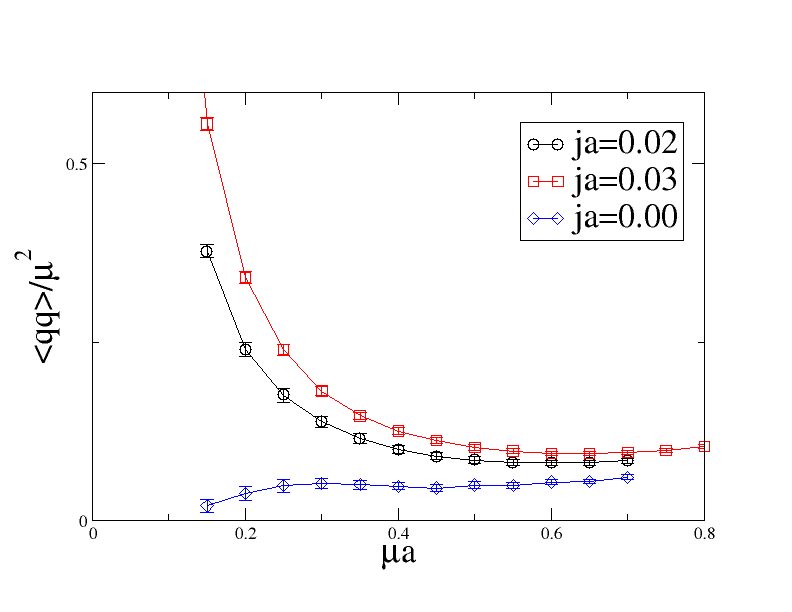}
\caption{Diquark condensates for $ja=0.02$ and $ja=0.03$ ($j=0$ extrapolated).}
\end{figure}

Figure 3 shows the quark number density with respect to chemical potential for $ja=0.02$ and $ja=0.03$. The curve for $j=0$, is extrapolated using linear extrapolation. The data are normalized by the continuum non-interacting quark number density, denoted by $n_{SB}$. We see a clear plateau in the region $\mu a = 0.3 - 0.7$, which roughly corresponds to $ n_{q} / n_{SB} = 1 $. This indicates that the system behaves like a non-interacting quark gas in this region.
\begin{figure}[h!]
\includegraphics[width=12.6cm,height=9.0cm]{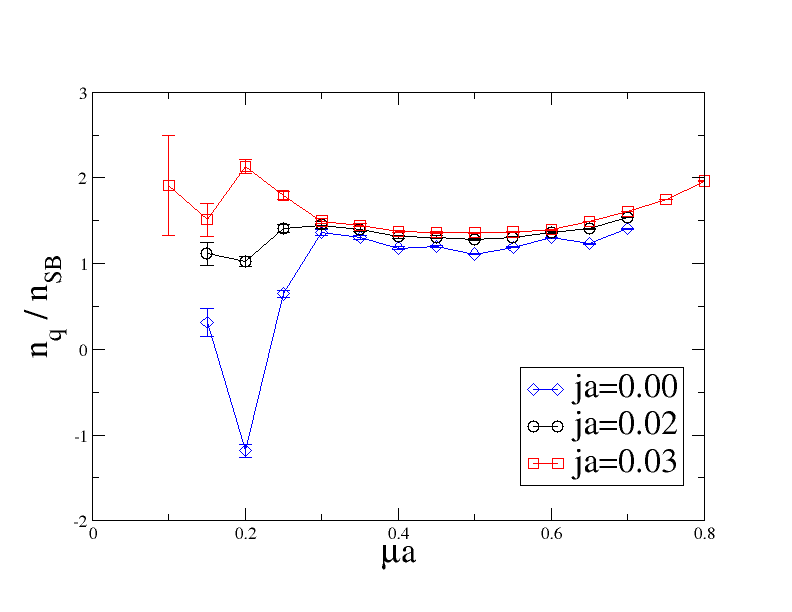}
\caption{Quark number density for $ja=0.02$ and $ja=0.03$ ($j=0$ extrapolated).}
\end{figure}

\section{The Gor'kov propagator and the form factors}
The action \eqref{eq.1} can be written in the compact form
\begin{equation}
S=\overline{\Psi}\mathcal{M}\Psi,
\end{equation}
where $\Psi\equiv\left(\begin{array}{c}\psi_{1}\\C^{-1}\tau_{2}\overline{\psi}_{2}^{tr}\end{array}\right)$, and 
\begin{equation}
\mathcal{M}=\left(\begin{array}{cc}
M\left(\mu\right) & -\frac{j}{2}C\gamma_{5}\tau_{2}\\
\frac{j}{2}C\gamma_{5}\tau_{2} & C\tau_{2}M\left(-\mu\right)C\tau_{2}
\end{array}\right)\equiv\left(\begin{array}{cc}
M & -A\\
A & \overline{M}
\end{array}\right).
\end{equation}

$\mathcal{M}$ is the Wilson fermion matrix in the presence of a non-zero diquark source and is known as the \textit{Gor'kov matrix}. The inverse of the Gor'kov matrix is the \textit{Gor'kov propagator}:

\begin{equation}
G=\mathcal{M}^{-1}\equiv\left(\begin{array}{cc}
S & T\\
\overline{T} & \overline{S}
\end{array}\right).
\end{equation}

The off-diagonal block components $T$ and $\overline{T}$ are responsible for the anomalous propagation of quarks, which turns a quark into an antiquark or vice-versa.

The $S$ and $T$ block components of the Gor'kov propagator can each be written in terms of four \textit{form factors}, which are useful tools to study the propagators:

\begin{equation}
S\left(p\right)
=\mbox{i}\mathcal{6}\mathbf{p}S_{a}\left(p\right)+S_{b}\left(p\right)+\mbox{i}\omega\gamma_{4}S_{c}
\left(p\right)+\mathcal{6}\mathbf{p}\gamma_{4}S_{d}\left(p\right),
\label{eq.8}
\end{equation}

\begin{equation}
T\left(p\right)=\mbox{i}\mathcal{6}\mathbf{p}T_{a}\left(p\right)+T_{b}\left(p\right)+\mbox{i}
\omega\gamma_{4}T_{c}\left(p\right)+\mathcal{6}\mathbf{p}\gamma_{4}T_{d}\left(p\right).
\label{eq.9}
\end{equation}

These expressions are in the continuum and are of the most general form possible which respects all symmetries of the theory \cite{Furnstahl1995}. For the expressions on the lattice we use the lattice momenta, $p=\mbox{sin}\left(pa\right) $, see \cite{Hands2000}.

Real and imaginary parts of the $S_{a}$ form factor corresponding to the normal propagation at various chemical potentials are given in Figure 4. The behaviour of both the real and the imaginary parts of the form factor changes as the chemical potential increases, which we interpret to be due to the normal to superfluid phase transition. Note that $S_{a}$ is not defined at $p_{s}=0$ according to \eqref{eq.8}.

\begin{figure}[h!]
\includegraphics[height=.28\textheight]{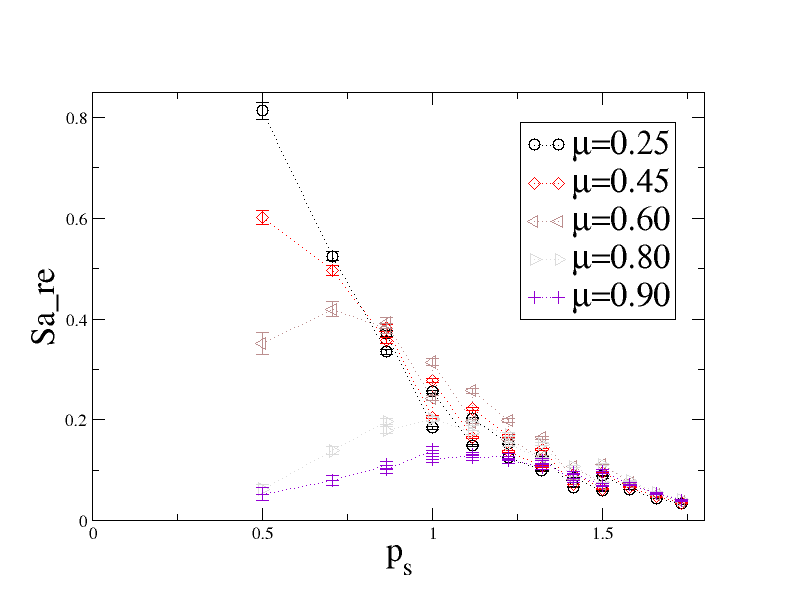}
\includegraphics[height=.28\textheight]{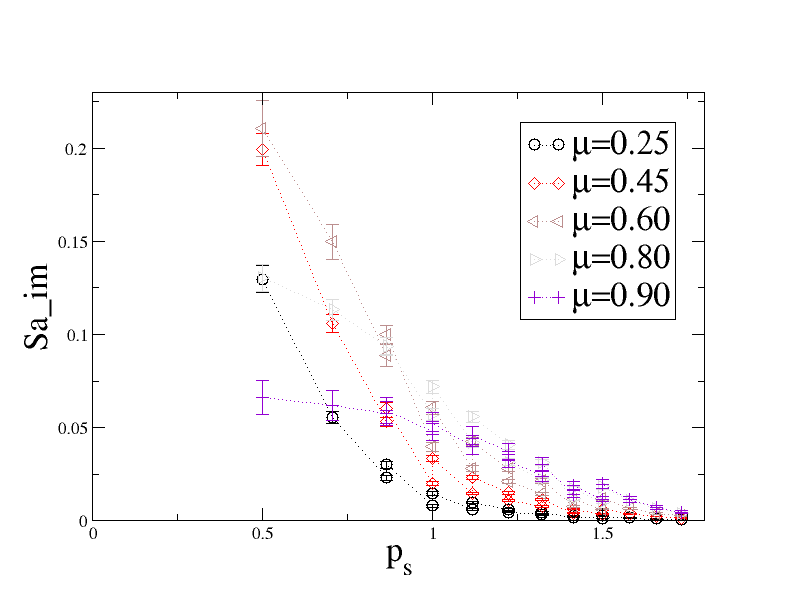}
\caption{Real and imaginary parts of the form factor $S_{a}$.}
\end{figure}

Real and imaginary parts of the $S_{b}$ form factor corresponding to the normal propagation at various chemical potentials are given in Figure 5. This form factor is related to the dynamical mass and is real. We see that the behavior of the real part of $S_{b}$ changes with increasing chemical potential, which is due to the change in the behaviour of the chiral condensate, hence, the dynamical mass. The imaginary part for high spatial momenta is consistent with zero, as expected. But there is a deviation from zero for low spatial momenta. According to \eqref{eq.3}, in the absence of a diquark source the form factor $S_{b}$ for free quarks is given by
\begin{equation}
S_{b}=\frac{m_{0}+\sum_{j=1}^{3}\left[\mathbf{1}-\mbox{cos}\left(p_{j}\right)\right]+\mathbf{1}-\mbox{cos}\left(\omega\right)}{\sum_{j=1}^{3}\mbox{sin}^{2}\left(p_{j}\right)+\mbox{sin}^{2}\left(\omega\right)+\left\{ m_{0}+\sum_{j=1}^{3}\left[\mathbf{1}-\mbox{cos}\left(p_{j}\right)\right]+\mathbf{1}-\mbox{cos}\left(\omega\right)\right\} ^{2}}.
\end{equation}
We see that as soon as $\mu\neq0$ an imaginary part will appear, which explains this deviation.
\begin{figure}[h!]
\includegraphics[height=.28\textheight]{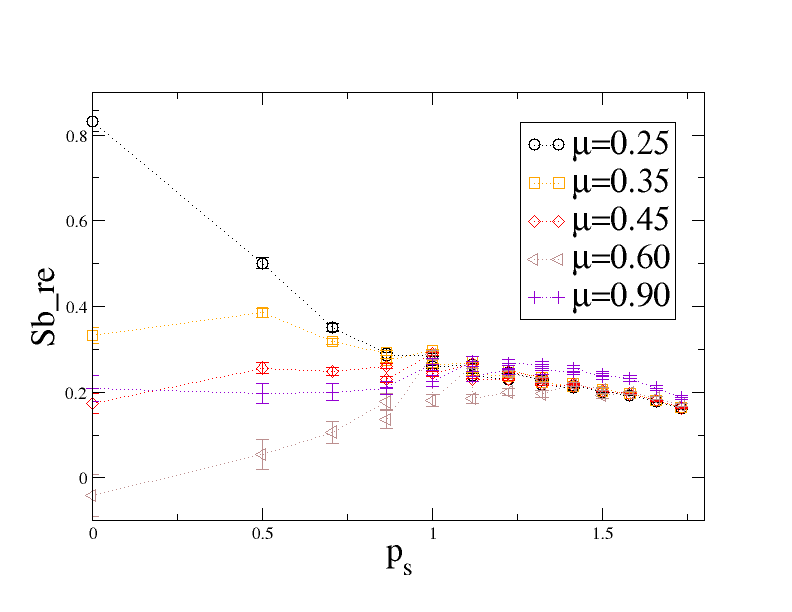}
\includegraphics[height=.28\textheight]{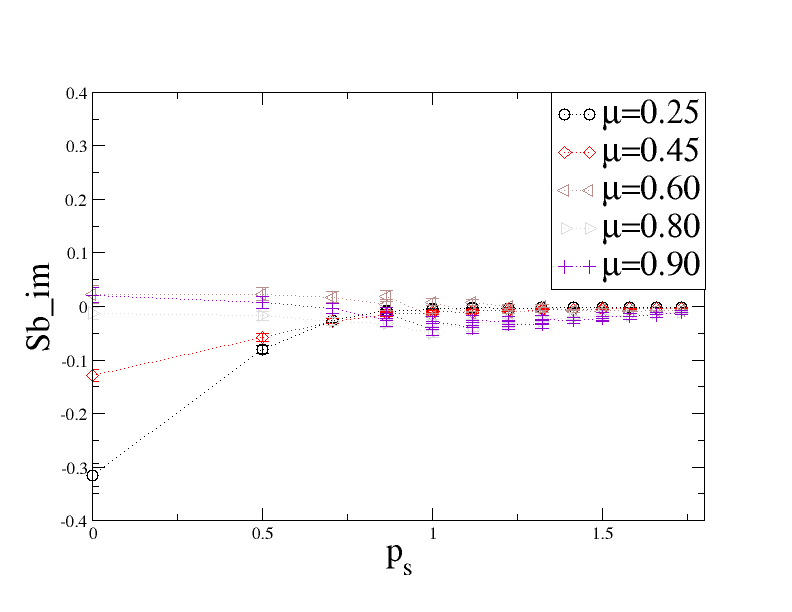}
\caption{Real and imaginary parts of the form factor $S_{b}$.}
\end{figure}

Real and imaginary parts of the $S_{c}$ form factor corresponding to the normal propagation at various chemical potentials are given in Figure 6. For low spatial momenta there is a decrease in the imaginary part with increasing chemical potential, while for high spatial momenta there is an increase. The location of the zero crossing is an indicator of a Fermi surface.

\begin{figure}[h!]
\includegraphics[height=.28\textheight]{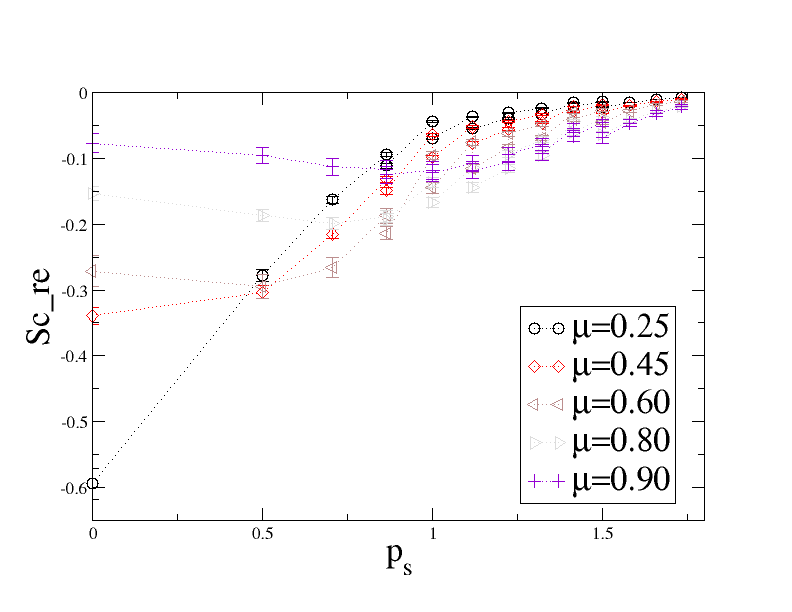}
\includegraphics[height=.28\textheight]{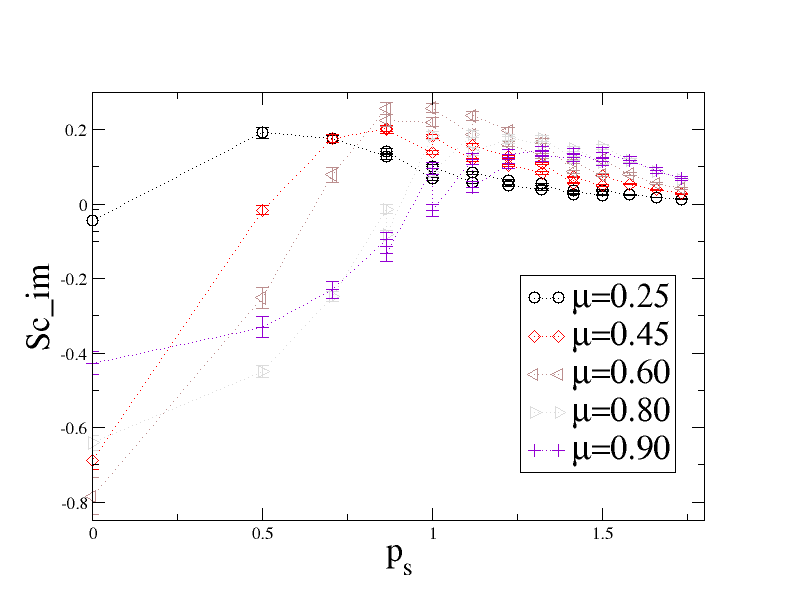}
\caption{Real and imaginary parts of the form factor $S_{c}$.}
\end{figure}

Real and imaginary parts of the form factor $S_{d}$ corresponding to the normal propagation and $T_{a}$ corresponding to the anomalous propagation have been found to be consistent with zero. The reason for $S_{d}$ to be zero is explained in \cite{Furnstahl1995}.

Real and imaginary parts of the $T_{b}$ form factor corresponding to the anomalous propagation at various chemical potentials are given in Figure 7. The imaginary part is consistent with zero while the real part is not. This is the form factor most directly associated with the diquark gap, which in most model studies is taken to be Dirac scalar.

\begin{figure}[h!]
\includegraphics[height=.28\textheight]{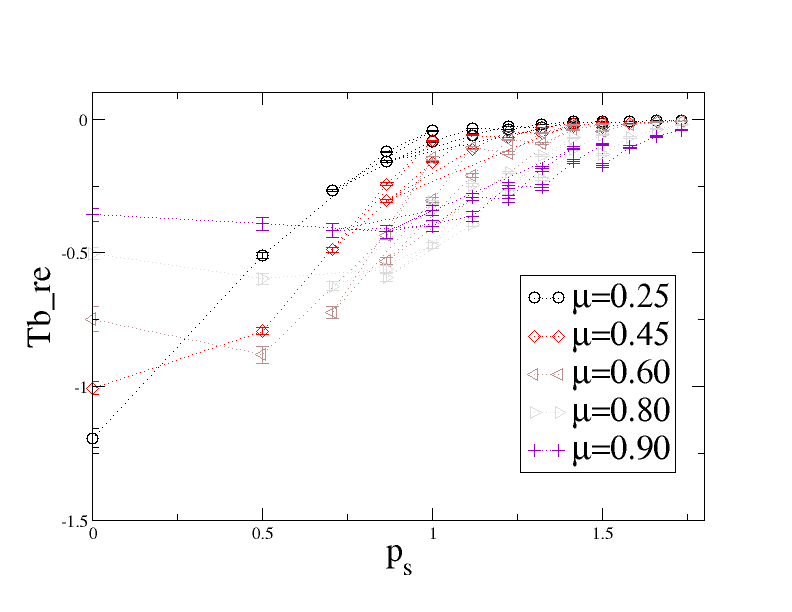}
\includegraphics[height=.28\textheight]{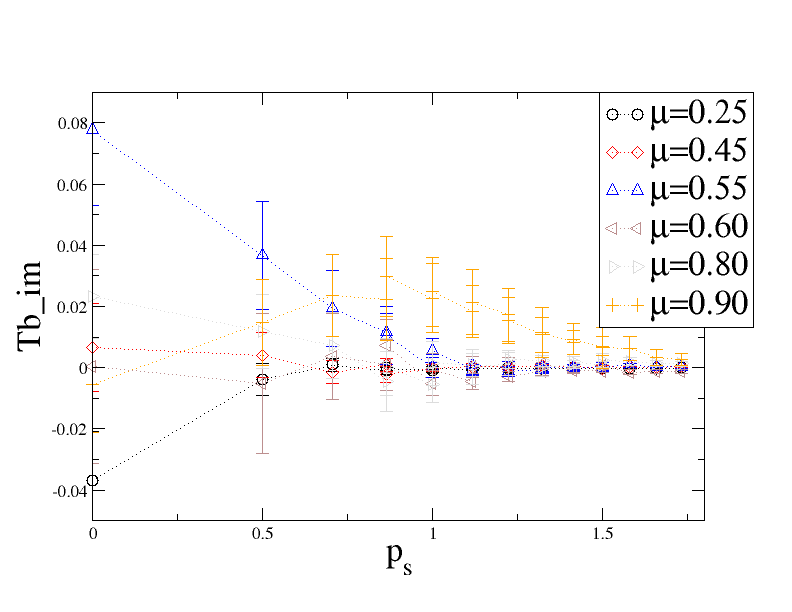}
\caption{Real and imaginary parts of the form factor $T_{b}$.}
\end{figure}

\pagebreak
Real and imaginary parts of the $T_{c}$ form factor corresponding to the anomalous propagation at various chemical potentials are given in Figure 8. While both the real and the imaginary parts are consistent with zero at high spatial momenta, there is a big deviation from zero at low chemical potentials. It might be that even at zero chemical potential, the nonzero diquark source induces diquark condensate, giving rise to this deviation observed.

\begin{figure}[h!t]
\includegraphics[height=.28\textheight]{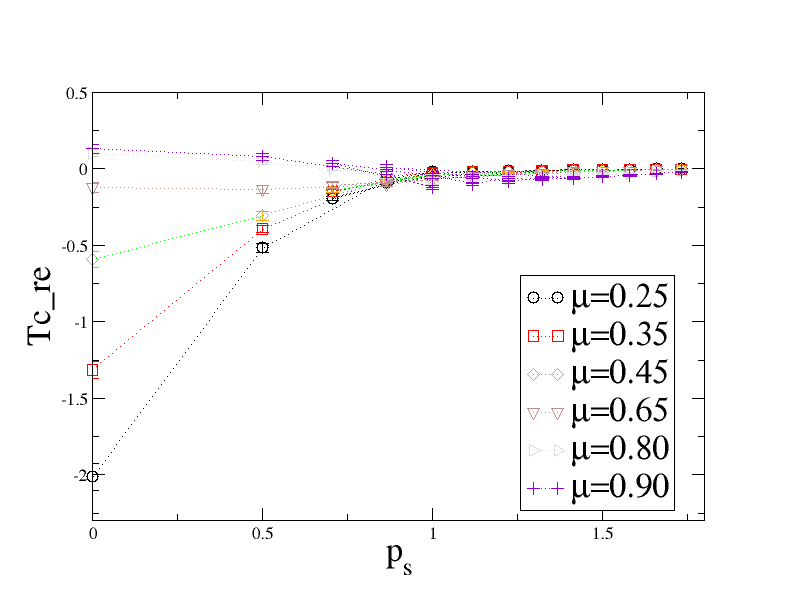}
\includegraphics[height=.28\textheight]{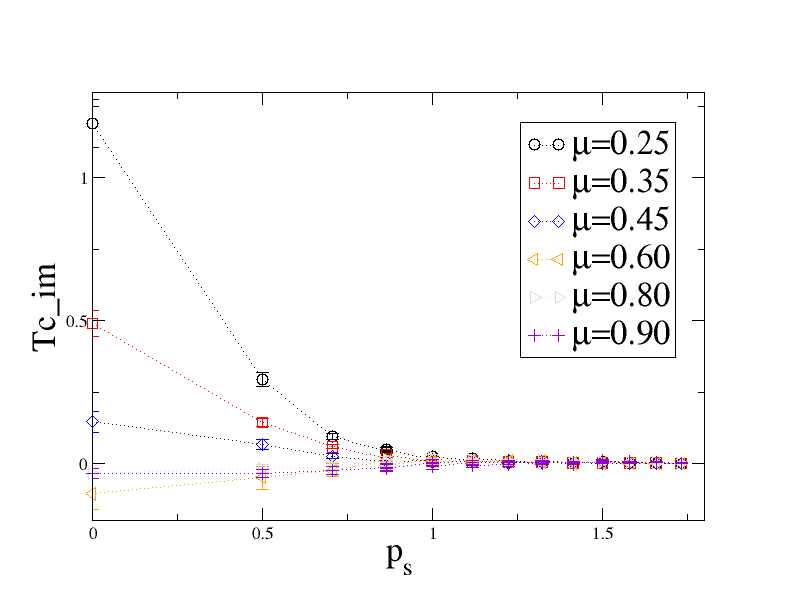}
\caption{Real and imaginary parts of the form factor $T_{c}$.}
\end{figure}

Real and imaginary parts of the $T_{d}$ form factor corresponding to the anomalous propagation at various chemical potentials are given in Figure 9. Unlike the $S_{d}$ form factor of the normal propagation, the $T_{d}$ form factor of the anomalous propagation is not consistent with zero. This suggests that the anomalous propagation cannot be modelled purely a Dirac scalar gap, but may also include a tensor component. Note that the argument in \cite{Furnstahl1995} for why $S_{d}=0$ does not hold for the anomalous propagator.
\begin{figure}[h!]
\includegraphics[height=.28\textheight]{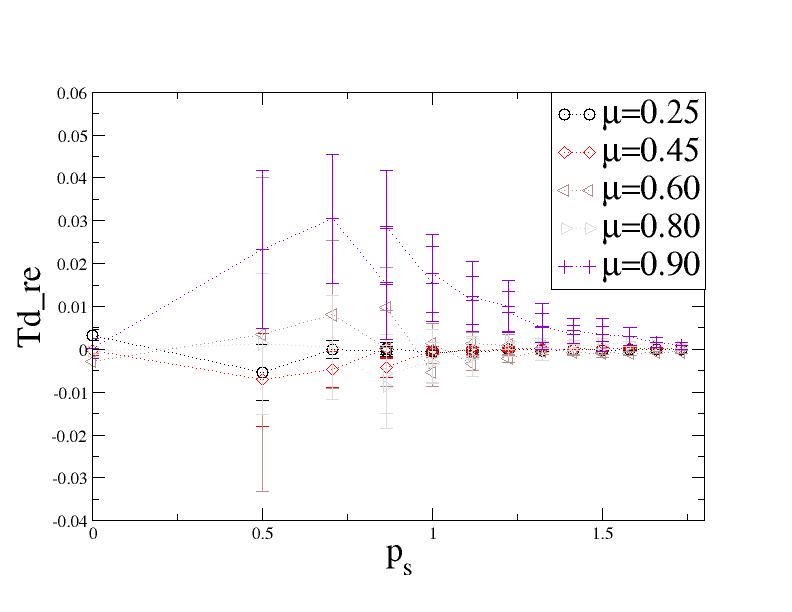}
\includegraphics[height=.28\textheight]{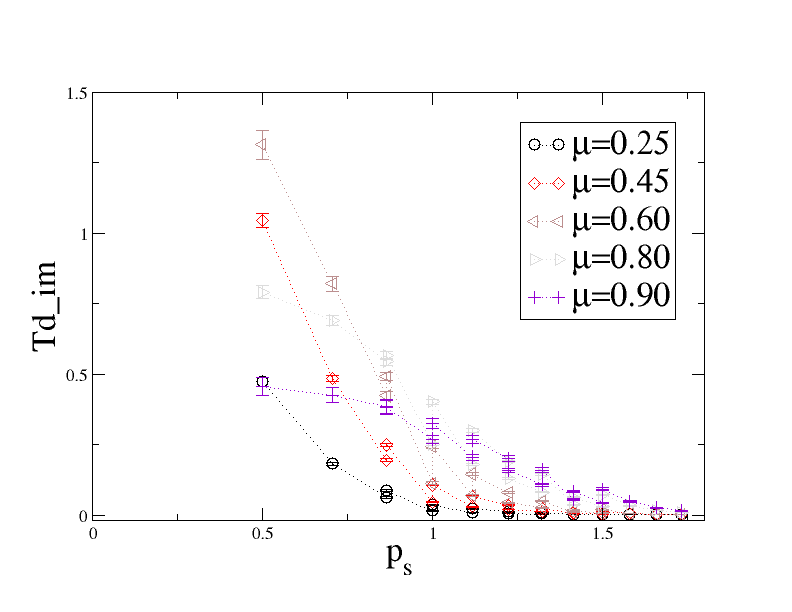}
\caption{Real and imaginary parts of the form factor $T_{d}$.}
\end{figure}

\pagebreak
\section{Summary and Outlook}
We have briefly discussed a tentative phase diagram of two-color QCD. We presented results for diquark condensate and found that a transition to a superfluid phase is observed at a chemical potential corresponding to half of the pion mass. This confirms and improves on previous results on coarser lattices. 

The diquark condensation was investigated using the diquark sources $ja=0.02$ and $ja=0.03$. In the future we want to do the same work including the data for $ja=0.01$.

We expressed the Gor'kov propagator in terms of the form factors and gave recent results from lattice simulations for them. 

The $S_{d}$ form factor for normal propagation was expected to be consistent with zero according to \cite{Furnstahl1995} and we have indeed observed this. Next we want to investigate analytically if the corresponding form factor for the anomalous propagator should be consistent with zero and see if our results from the simulation are reasonable. We also plan to compute the form factors of the inverse Gor'kov propagator.

\begin{theacknowledgments}
This work has been carried out with the support of Science Foundation Ireland. We acknowledge the use of the computational resources provided by the UKQCD collaboration and the DiRAC Facility jointly funded by STFC, the Large Facilities Capital Fund of BIS and Swansea University. Tamer Boz and Jon-Ivar Skullerud thank University of Adelaide for their hospitality.
\end{theacknowledgments}

\end{document}